%% file: main.tex
\documentclass[useAMS,usenatbib]{mnras} 
\usepackage{graphicx}
\usepackage{amsmath}
\usepackage{bm}
\usepackage{url}
\usepackage{color}
\usepackage{arydshln}
\usepackage{lscape}
\usepackage[section]{placeins}
\usepackage[normalem]{ulem}
\usepackage{lineno}

\begin{document}


\title{Gravitational redshift profiles of MaNGA BCGs}

\author[Zhu et al.]{\parbox{18cm}{Hongyu Zhu$^{1,2}$\thanks{E-mail:
hongyuz@andrew.cmu.edu}, Shadab Alam$^{3,1,2}$, Rupert A. C. Croft$^{1,2,4}$, Shirley Ho$^{1,2,5,6,7}$, Elena Giusarma$^{1,2,5,6}$, Alexie Leauthaud$^{8,9}$, and Michael Merrifield$^{10}$}\vspace{0.3cm}\\
    $^{1}$ Department of Physics, Carnegie Mellon University, 5000 Forbes Ave., Pittsburgh, PA 15213 \\
    $^{2}$ McWilliams Center for Cosmology, Carnegie Mellon University, 5000 Forbes Ave., Pittsburgh, PA 15213 \\
    $^{3}$ Institute for Astronomy, University of Edinburgh, Royal Observatory, Blackford Hill, Edinburgh, EH9 3HJ , UK\\
    $^{4}$ ASTRO-3D Center, School of Physics, University of Melbourne, Parkville, VIC 3010, Australia\\
    $^{5}$ Berkeley Center for Cosmological Physics, University of California, Berkeley, CA 94720, USA\\
    $^{6}$ Lawrence Berkeley National Laboratory (LBNL), Physics Division, Berkeley, CA 94720, USA\\
    $^{7}$ Flatiron Institute, Center for Computational Astrophysics, 162 Fifth Ave., New York, NY, USA 10010\\
    $^{8}$ Kavli IPMU (WPI), UTIAS, The University of Tokyo, Kashiwa, Chiba 277-8583, Japan\\
    $^{9}$ Department of Astronomy and Astrophysics, UCO/Lick Observatory, University of California, 1156 High Street, Santa Cruz,
    CA 95064, USA\\
    $^{10}$ School of Physics and Astronomy, University of Nottingham, University Park, Nottingham NG7 2RD, UK\\ 
}
    
\date{\today}
\pagerange{\pageref{firstpage}--\pageref{lastpage}}   \pubyear{2018}
\maketitle
\label{firstpage}

\input{tex/abstract}

\begin{keywords}
gravity, dark matter
\end{keywords}

\input{tex/intro}

\input{tex/theory}

\input{tex/data}

\input{tex/dap}

\input{tex/dsp}

\input{tex/discussion}

\input{tex/conclusion}

\section*{Acknowledgments}
This work was supported by NSF Award AST-1412966. SA is also
supported by the European Research Council through the COSFORM
Research Grant (\#670193). 

Funding for the Sloan Digital Sky Survey IV has been provided by the Alfred P. Sloan Foundation, the U.S. Department of Energy Office of Science, and the Participating Institutions. SDSS acknowledges support and resources from the Center for High-Performance Computing at the University of Utah. The SDSS web site is \href{www.sdss.org}{www.sdss.org}.

SDSS is managed by the Astrophysical Research Consortium for the Participating Institutions of the SDSS Collaboration including the Brazilian Participation Group, the Carnegie Institution for Science, Carnegie Mellon University, the Chilean Participation Group, the French Participation Group, Harvard-Smithsonian Center for Astrophysics, Instituto de Astrofísica de Canarias, The Johns Hopkins University, Kavli Institute for the Physics and Mathematics of the Universe (IPMU) / University of Tokyo, the Korean Participation Group, Lawrence Berkeley National Laboratory, Leibniz Institut für Astrophysik Potsdam (AIP), Max-Planck-Institut für Astronomie (MPIA Heidelberg), Max-Planck-Institut für Astrophysik (MPA Garching), Max-Planck-Institut für Extraterrestrische Physik (MPE), National Astronomical Observatories of China, New Mexico State University, New York University, University of Notre Dame, Observatório Nacional / MCTI, The Ohio State University, Pennsylvania State University, Shanghai Astronomical Observatory, United Kingdom Participation Group, Universidad Nacional Autónoma de México, University of Arizona, University of Colorado Boulder, University of Oxford, University of Portsmouth, University of Utah, University of Virginia, University of Washington, University of Wisconsin, Vanderbilt University, and Yale University.


\bibliography{Master_Hongyu.bib}
\bibliographystyle{mnras}
\input{tex/appendix}

\label{lastpage}

\end{document}

%% file: tex/abstract.tex
\begin{abstract}
The gravitational potential well of an $M>10^{13}$ $M_\odot$ galaxy will lead to a gravitational redshift differential of order 1 km/s between the galaxy core and its outskirts. Current surveys of massive galaxies with spatially resolved spectroscopy have reached a size which makes feasible attempts to detect gravitational redshifts within galaxies. We use spectra from the  Mapping Nearby Galaxies at Apache Point Observatory (MaNGA) experiment to attempt a measurement of the averaged stellar redshift profile of large elliptical galaxies.  We find that systematic effects (possibly related to charge transfer or wavelength calibration accuracy) make the standard MaNGA data pipeline unsuitable for measuring the relevant sub km/s wavelength shifts. We therefore develop a cross-correlation technique to mitigate these effects, but caution that we are working beyond the design accuracy of the MaNGA experiment. With a sample of $272$ galaxies in halos with $\log (M/M_\odot)>13$, we attempt a measurement of the gravitational redshift profile, achieving $1 \sigma$  errors of size $\sim 0.5$ km/s,  but are unable to  make a significant detection of the signal. Even without a detection, our  measurement can be used to limit the dark matter mass within the half light radius of elliptical galaxies to $1.2$ times the stellar mass, at the 68\% confidence level. We also perform weighting scheme tests and split sample tests, and address 
target selection issues and other relativistic effects, including the transverse Doppler effect and relativistic beaming of stars. Future detections could lead to new constraints on the galaxy mass distribution that are different from kinematic and lensing determinations and open a window on galaxy properties and tests of gravity.
\end{abstract}

%% file: tex/intro.tex
\section{Introduction} \label{sec:intro}
The gravitational redshift is one of the fundamental predictions of Einstein's General Relativity \cite{Einstein1916} and has long been considered as a component of the total redshift of galaxies \citep{pound1959, greenstein1971, lopresto1991}. It can be used to test our theoretical understanding, and probe certain unique aspects of cosmic structures. The gravitational redshift is the relative shift in wavelength of light as it moves within the gravitational field of massive objects. This shift is typically very small but can become significant in cosmological scales around massive dark matter halos and galaxies. In the weak field limit, the gravitational redshift (denoted as $z_g$) of photons with wavelength $\lambda$ is $z_g=\Delta\lambda/\lambda=\Delta\phi/c^2$ where $\phi$ is the gravitational potential at the point where photon was emitted. Theoretically, we can predict the gravitational redshift experienced by a photon emitted from a galaxy as it travels out of the gravitational potential of the galaxy and bigger dark matter halos surrounding such a galaxy. \cite{cappi1995} predicted the gravitational redshift difference to be $\sim 10-100$ km/s between the centers and outskirts of massive galaxy clusters assuming the commonly used de Vaucouleurs mass distribution (\citealt{de1948}). Such a signal of the gravitational redshift will be hidden under a few orders of magnitude larger shift due to the peculiar velocities of the galaxies and a further shift due to cosmological Hubble flow. Therefore, it will be impossible to detect such signal for individual galaxies with current observational precision, but by averaging over the noise using many clusters it is possible to make a statistical measurement, as was shown by \cite{kim2004} who made predictions for the amplitude using galaxy clusters in an $N$-body simulations of $\Lambda$CDM universe. Such a statistical detection of gravitational redshifts from clusters necessitates  large galaxy redshift surveys. This is also the case for a measurement on larger scales, as by measuring the amplitude of the asymmetry on 2-d cross-correlation function, \cite{croft2013} showed that a $\sim 4\sigma$ detection should be expected from the full Baryonic Oscillation Spectroscopic Survey (BOSS). Measurements with precision at the few per cent level may be expected from upcoming larger redshift surveys. However, the gravitational redshift is not the only effect that causes the asymmetry, other relativistic effects such as the transverse Doppler effect \citep{kaiser2013, zhao2013, Giusarma2016PT}, luminosity distance perturbation \citep{kaiser2015, Giusarma2016PT} and special relativistic beaming \citep{kaiser2015, Alam2016TS, Giusarma2016PT} can also contribute to this asymmetry. \cite{Zhu2016Nbody} modelled these relativistic effects in the simulation and found the gravitational redshift dominated over other effects in the non-linear regime. Corrections were also proposed by \cite{Zhu2016Nbody, cai2016}. The non-linear regime is further explored in the hydrodynamical simulation \citep{Zhu2016Hydro}[in prep.]. The asymmetry on the cross-correlation function has also been studied extensively from using perturbation theory by e.g., \cite{bonvin2014, yoo2009, yoo2012, Breton2018}.

The gravitational redshift is so small compared to for example the redshift caused by peculiar velocities that it has become feasible to measure in cosmological observations only quite recently. The first cosmological observational measurements of gravitational redshifts in galaxy clusters were carried out by \cite{wojtak2011}, using 7800 clusters from the SDSS survey. Other measurements at similar significance level were made by \cite{dominguez2012}, \cite{jimeno2015} and also \cite{sadeh2015}. Including the other relativistic effects mentioned above, \cite{Alam2016Measurement} also measured the gravitational redshift signal on larger scales using galaxy cross-correlation functions.

On smaller, galactic scales, there should be a gravitational redshift
difference between the centers and outskirts of galaxies. For the most massive galaxies, this could be a
fraction of a km/s or more. A first attempt to measure this, from individual galaxies was carried out
by \cite{Stiavelli1993}, but a more recent study by \cite{coggins2003} showed that this
earlier work was unlikely to have detected a signal. \cite{coggins2003} also showed that even in massive galaxies detection is not possible on an individual basis, but that extremely compact 
dwarf galaxies have steep enough gravitational redshift profiles to make them promising candidates.
In this paper, we apply a similar statistical approach to that successfully employed by \cite{wojtak2011}, but this time on galactic rather than cluster scales. We attempt to make a detection of gravitational redshift from stacked MaNGA \citep{bundy2014} Brightest Cluster Galaxies (BCGs), along with consideration of other effects \citep{zhao2013, kaiser2013, Zhu2016Nbody, Giusarma2016PT}. The major difficulty in detecting gravitational redshifts in galaxies is that the signal is  one to two orders of magnitude weaker than that in galaxy clusters.

Alongside making a detection, we need a simple but accurate theoretical prediction for gravitational redshifts inside galaxies. The de-Vaucouleurs law \citep{de1948} and its generalization the Sersic law \citep{sersic1963} fit the brightness profiles of elliptical galaxy bulges remarkably well. However, these are projected quantities. Further studies such as dynamical studies of elliptical galaxies involve the deprojection of observed (projected) quantities, which are used to derive other spatial quantities. For instance, in this paper, the projected density profile follows the de-Vaucouleurs profile under the assumption that light traces mass \citep{West1989}, however the spatial (deprojected) density distribution is more useful for studying the spatial potential profile. Many attempts have been made to transform the projected de Vaucouleurs law or Sersic law into the corresponding spatial profile. Some results are given in the form of numerical tables (e.g. \citealt{Poveda1960, Young1976}) and some are approximations and asymptotic expressions (e.g. \citealt{Jaffe1983, Mellier1987, hernquist1990, Ciotti1991, Graham1997, Marquez2001}). \cite{Jaffe1983} and \cite{hernquist1990} models are widely used, however neither the Jaffe nor Hernquist profiles are able to approximate the inner part well. Therefore, an extra parametrization is needed. The Dehnnen model \citep{Dehnen1993}, also called the``gamma profile'' generalizes both \cite{Jaffe1983} and \cite{hernquist1990} by parametrizing the central slopes.
\begin{equation}\label{eq:jhmodels}
\rho(r) = \frac {(3-\gamma)M} {4\pi}\frac {a}{r^\gamma (r+a)^{4-\gamma}},
\end{equation}
where $M$ is the total mass and $a$ a scale radius. $\gamma$ is restricted to the interval $[0, 3]$. The models by Jaffe and Hernquist correspond to $\gamma=2$ and $\gamma=1$, leading to the slopes of the mass distribution in the center being $r^{-2}$ and $r^{-1}$ respectively. At a sufficiently large radius, $\rho(r) \propto r^{-4}$.  This model can easily be extended to a more general form $\rho(r)\propto r^{-\gamma}(1+r^{1/\alpha})^{(\gamma-\beta)\alpha}$ where $(\alpha, \beta, \gamma) = (1,4,2), (1,4,1), (1,3,1), (1,4,1.5), (0.5,5,0)$ represent Jaffe, Hernquist, NFW \citep{navarro1996, navarro1997}, \cite{moore1999}, and \cite{Plummer1911} profiles respectively.

Our plan for the paper is as follows. In Sec.~\ref{sec:theory} we find the theoretical form for the gravitational potential, and hence redshift profile inside galaxies as well as the transverse Doppler effect. In Sec.~\ref{sec:data} we introduce two MaNGA products and explain how we select large ellipticals for our samples. We then perform analysis on both MaNGA products in Sec.~\ref{sec:dap} and Sec.~\ref{sec:dsp} respectively. We conclude in Sec.~\ref{sec:conclusion} with a summary of our findings.

%% file: tex/theory.tex
\section{Theoretical redshift profile} \label{sec:theory}

\subsection{Gravitational redshift}

\cite{cappi1995} assumed that galaxy clusters obey a de Vaucouleurs profile and estimated their expected gravitational redshift profiles. In that paper a Monte Carlo technique was adopted with simulating points, each being proportional to the mass contained in each spherical bin for each fixed $M$ and $R_e$ and averaging the points in the bin weighted by the potentials. The mass and the potentials were calculated using the \citep{Young1976} values. In Fig.~1 of \cite{cappi1995}, the author shows that the Hernquist model underestimates the spatial potential with respect to the de Vaucouleurs law at small scales, using the integral form of Poisson Equation: Eqn.~\ref{eq:poisson}. Here we use the same technique, taking the density distribution Eqn.~\ref{eq:jhmodels}, applying the Poisson Equation: $\nabla ^2 \Phi_\mathrm{3d}(\mathbf{r}) = 4 \pi G\rho(\mathbf{r})$, and computing the spatial potential profile. We find that the Jaffe model overestimates the spatial potential compared to the de Vaucouleurs law. In order to find a good approximation to the potential profile on small scales, we allow the parameter $\gamma$ in Eqn.~\ref{eq:jhmodels} to vary, with $\gamma\in(1,2)$. This allows us to match the region where $r<0.2R_e$. Performing a simple least square fit to the de Vaucouleurs profile \citep{de1948}, we find the best fit $\gamma$ to be 1.24.

\begin{equation}\label{eq:poisson}
\Phi_\mathrm{3d}(r) = -4\pi G\left[\frac 1 r \int\limits_0^r \rho(x) x^2 \mathrm{d}x + \int\limits_r^\infty \rho(x) x \mathrm{d}x\right].
\end{equation}

To convert the spatial potential to observable quantities, we need to project $\Phi_\mathrm{3d}$ onto a plane by sampling the points from the density distribution, $\rho(r)$ and averaging the $\Phi_\mathrm{3d}$, or equivalently the line-of-sight velocities along the $z$-axis. We recall that telescopes collect photons emitted from any position in a manner proportional to  the line-of-sight stellar luminosity density. The projection of  luminosity weighted velocities of these stars serves to average the velocities. Therefore, the following transformation is performed:
\begin{equation}
\Phi_\mathrm{2d}(R) = \frac {\int\limits_0^\infty\Phi_\mathrm{3d}(R,z)\rho(R,z)\mathrm{d}z}{\int\limits_0^\infty\rho(R,z)\mathrm{d}z} = \frac {\int\limits_R^\infty\Phi_\mathrm{3d}(r)\rho(r)r/\sqrt{r^2-R^2}\mathrm{d}r}{\int\limits_R^\infty\rho(r)r/\sqrt{r^2-R^2}\mathrm{d}r},
\end{equation}
where $R$ denotes the distance to the center in the plane. The final equation used to compute $\Phi_\mathrm{2d}(R)$ is then:
\begin{equation}\label{eq:phi}
\begin{split}
\Phi_\mathrm{2d}(R)&=-4\pi G\frac{\int\limits_R^\infty \mathrm{d}r\rho(r)/\sqrt{r^2-R^2}\int\limits_0^r \mathrm{d}x\rho(x)x^2}{\int\limits_R^\infty\rho(r)r/\sqrt{r^2-R^2}\mathrm{d}r}\\
&+\frac{\int\limits_R^\infty \mathrm{d}r\rho(r)r/\sqrt{r^2-R^2}\int\limits_r^\infty \mathrm{d}x\rho(x)x} {\int\limits_R^\infty\rho(r)r/\sqrt{r^2-R^2}\mathrm{d}r}.
\end{split}
\end{equation}

\subsection{Transverse Doppler and other effects}

The transverse Doppler effect is the change in the observed wavelength of light compared to emitted wavelength when the source moves perpendicular to the observer's line of sight. It is due to time dilation rather than the usual Doppler effect. Though the radial averaging of the velocity profile reduces the effect of noise from e.g., substructures and inner rotations, it is not yet a pure gravitational redshift profile but a combination of at least three different effects, the gravitational redshift, the transverse Doppler effect and the special relativistic beaming effect. 

We find that the effect of special relativistic beaming is very small compared to the other two. We evaluate its influence following the methodology described in Sec. 2.4 in \cite{Zhu2016Nbody}. We apply weights which are  a function of the line-of-sight velocities of stars in galaxies (Eqn.~7 in \citealt{Zhu2016Nbody}). The weights depend on the spectral energy distribution of the stellar population, and values used in the literature range from  $f_\text{beam}=1$ used by \cite{Zhu2016Nbody} for LRG BOSS galaxies to $f_\text{beam}=6$ used by \cite{kaiser2013} for SDSS cluster galaxies. 
 We find the change in the $z_g$ signal on all scales is less than 1\% even with much larger $f_\text{beam}$ values ($f_\text{beam}=100$).  We therefore neglect special relativistic beaming of stars in the rest of our analysis. 
 
 In order to take into account the transverse Doppler effect, in Sec.~\ref{sec:dap} and Sec.~\ref{sec:dsp} we  fit the theoretical curve $\left(\Phi-\mathbf{v}^2/2\right)/c$ to the radial profile, where $\Phi$ is the projected potential profile from the Dehnnen model and $-\mathbf{v}^2/2c$ accounts for the transverse Doppler effect. Here $\mathbf{v}$ is the 3d velocity. For each spaxel (a.k.a spatial pixel),
\begin{equation}
|\mathbf{v}_i|^2 = v_{i,\mathrm{los}}^2 + \sigma_i^2 - \sigma_{i,\mathrm{corr}}^2,
\end{equation}
where $\sigma_{i,\mathrm{corr}}$ accounts for the spaxel-wise correction of the instrumental resolution effects.

%% file: tex/data.tex
\section{Data} \label{sec:data}

\begin{figure*}
\centering
\includegraphics{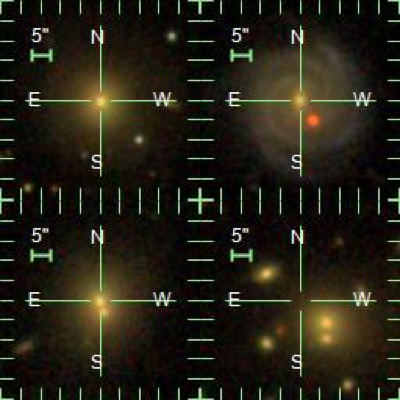}
\caption{Four sample galaxy images in our MaNGA sample with \texttt{GRP\_MHALO\_LEST>13}. The top panels show good detections of ellipticals (left, plateID: 7962, designID: 6103) and spirals (right, plateID: 8459, designID: 3703) while the bottom panels are considered  bad detections. The bottom left panel (plateID: 8554, designID: 3702) shows two extremely close galaxies and we believe the potential profile of the central galaxy would be highly contaminated by the other one. The bottom right panel (plateID: 8601, designID: 12704) shows no galaxy in the center. We only include the well-detected ellipticals in our final sample.}
\label{fig:manga}
\end{figure*}

\begin{figure*}
\centering
\includegraphics[width=.8\textwidth]{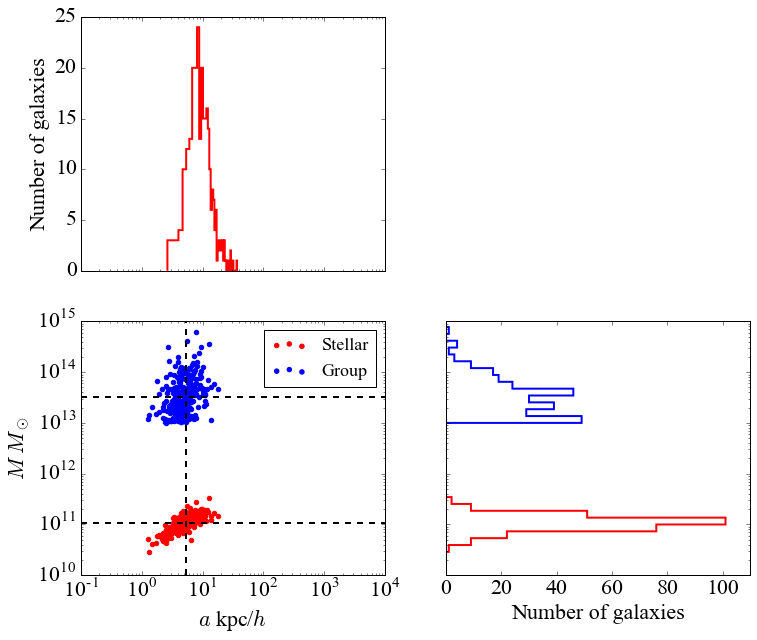}
\caption{Scatter plot and histograms of stellar mass, group mass and scale radius of sample galaxies. On the bottom left panel, the red and blue dots represent stellar and group mass respectively with scale radius of sample galaxies. The vertical black dashed line is the mean sample scale radius $a$ and the horizontal black dashed lines are sample group mass (top) and stellar mass (bottom). The top and right panel show the histograms of scale radius and masses, keeping the corresponding axis scales the same.}
\label{fig:sample_stats}
\end{figure*}

The Sloan Digital Sky Survey (SDSS, \citealt{eisenstein2011, Blanton2017}) has measured spectra for more than three million astronomical objects across one third of the sky. Spatially resolved observations such as the Mapping Nearby Galaxies at Apache Point Observatory (MaNGA, \citealt{bundy2014, Yan2016, Wake2017}) survey with integral field units (IFUs, \citealt{Drory2015, Law2015, Yan20161}) enables us to resolve the internal structure and velocities of $\sim 10,000$ nearby galaxies. The pipeline errors in MaNGA individual fiber velocities are $\sim 20$ km/s \citep{Yan2016} and by stacking enough large galaxies we anticipate that a statistically significant detection of relativistic effects may be possible.

\subsection{MaNGA products}

The MaNGA data products have been released approximately every year in \textit{MaNGA Product Launches} (MPLs), including different types of data with various levels of reliability. In this paper we use MPL-6, released in the summer of 2017 \citep{Albereti2017}. The products include those from the Data Reduction Pipeline (DRP) and those from the Data Analysis Pipeline (DAP). The DAP takes raw data files (arcs, flats, science frames), and produces extracted, sky-subtracted, flux calibrated individual exposures for each plate and further reduces to row-stacked-spectra (RSS) and data cubes for each target \citep{Law2016}. The DAP provides scientific measurements made from the data such as stellar continuum fits, emission line flux maps, and velocity maps.

\subsection{BCG samples}

We are interested in BCGs or larger elliptical galaxies since they are expected to have stronger gravitational redshift signals and relatively stable and well understood mass and light profiles. We use the MaNGA catalog merged with Yang et al. to select BCGs. Our criteria are \texttt{DRP\_MATCH==1, GRP\_MHALO\_LEST>13, BCG\_VIS==1, VIS\_BCG\_FLAG!=0, VIS\_CLUSTER\_FLAG==1,2,5}. This results in 383 galaxies, and we eyeball the resulting galaxy images to select 272 of the sample which are suitable ellipticals. We reject spirals (top right panel in Fig.~\ref{fig:manga}), galaxies with very close satellites (bottom left panel in Fig.~\ref{fig:manga}) and galaxies which are probably mis-centered (bottom right panel in Fig.~\ref{fig:manga}).



After picking these ellipticals, we plot the sample statistics related to mass and scale radius  in Fig.~\ref{fig:sample_stats}. The scale radii are computed based on the $\gamma=1.24$ Dehnnen model (Eqn.~\ref{eq:jhmodels}), for which $a=R_e/2.07$. The means and standard deviations of the logarithm of the  group, stellar mass and scale radius are $\log_{10}(M/M_\odot)=13.48\pm0.36, 11.03\pm0.15$ and $a=5.30\pm2.48$ (equivalent to a half light radius $R_e=11.0\pm5.13$) respectively.

%% file: tex/dap.tex
\section{Analysis of MaNGA DAP outputs} \label{sec:dap}

As mentioned in Sec.~\ref{sec:data}, the standard MaNGA data pipeline provides many data products
involving various levels of processing. For our purposes, the most straightforward analysis involves taking the velocities in galaxy spaxels provided by the MaNGA DAP, stacking them, and averaging them in radial bins to provide a redshift profile. We describe this procedure and
our results in this section. The spaxel velocities have been estimated by the pipeline to
be accurate to $\sim$ 20 km/s \citep{bundy2014,Yan2016}.
For our purposes, however we require accuracy at the sub km/s level, and so it has been
necessary to use the data closer to the raw form (the MaNGA DRP output). This analysis is described in Sec.~\ref{sec:dsp}. 

\subsection{Data mask}
\begin{figure*}
\centering
\includegraphics[width=1 \textwidth]{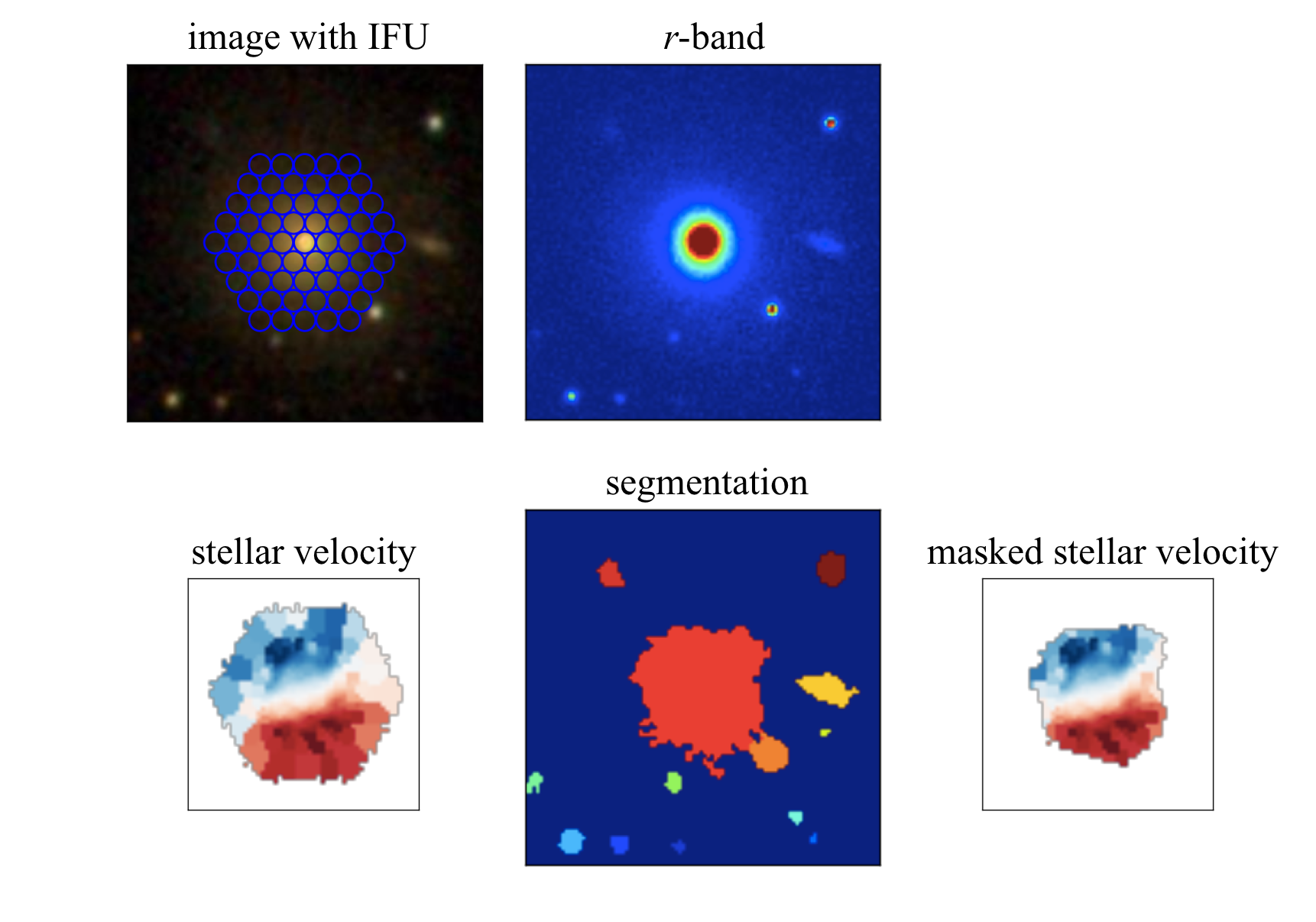}
\caption{The IFU geometry image (top left), $r$-band image (top right), stellar velocity map (bottom left), segmentation map (bottom middle) and the masked velocity map (bottom right) for one of well-detected MaNGA ellipticals (plateID: 7962, designID: 6103) are depicted. The top left panel shows the full-band image with 61 IFUs, which generates the output velocity map on the bottom left, where blue indicates blueshift and red indicates redshift. We can see that this elliptical has moderate inner rotations. However the inner rotations are averaged out on radial profiles. The top right panel shows the $r$-band image and this was used to generate the segmentation map in the bottom middle. The bottom right panel shows the pipeline output with satellite galaxy information masked. This process is done for each  of the 272 ellipticals in our sample.}
\label{fig:seg}
\end{figure*}

We generate the  BCG sample following the procedure described in Sec.~\ref{sec:data}. For each velocity map, we compute the mean velocity within 3 arcsecs of the galaxy center, and subtract this from the velocity in each spaxel. We also take care of the extreme values and outliers by applying a pixel quality S/N mask (\texttt{S/N>8}). Moreover we expect that the signals could be contaminated by close satellites, and for this reason we decide to remove them from our analysis. Fig.~\ref{fig:seg} shows the process we go through to eliminate satellites. Starting with a good elliptical detection with 61 IFUs, we query and download $r$-band fits images from the SDSS server, run source extractors \citep{Bertin1996} and create a satellite mask (segmentation map). However, satellite masks do not always have complete coverage in azimuth (as illustrated in Fig.~\ref{fig:seg}). This could lead to issues when a galaxy shows some net rotation (as in this example), as there will be a net velocity due to the incomplete sampling in azimuth even if it would  average to zero with full sampling. On taking the mean over many galaxies this effect will average out, but it could be a significant source of unnecessary noise in the data. We eliminate such net rotations by using an additional mask, a "mirror mask". The mirror mask is defined as follows: We  only include the data from a pixel with coordinated ($x$, $y$) where there is also data for the same galaxy for the pixel with coordinates (-$x$, -$y$). This masking of the data should cancel out the contribution from any reasonably symmetric velocity field.

\subsection{Stacked redshift profile of galaxies}

We convert velocity maps of our selected 272 galaxies from arcsecs to their physical scales in kpc/$h$. We then stack these maps by finding the medians of the velocities in radial bins from 0 to 30 kpc, with the error bars estimated from bootstrapping the whole sample for 5000 iterations. More specifically, each bootstrapping sample contains 272 galaxies randomly sampled with replacement. At the $k$-th iteration, the gravitational redshift is computed as $\mathbf{z}_k=(z_{1k},...,z_{nk})^\mathrm{T}$ where $n$ is the number of bins and $z_{ij}$ denotes the median over the redshifts of all the pixels falling into the $i$-th bin. The entry $\bm{\Sigma}_{ij}$ of the empirical covariance matrix is estimated using
\begin{equation}
\bm{\Sigma}_{ij} = \frac 1 B \sum\limits_{k=1}^{B} \left({z}_{ik}-\frac 1 B \sum\limits_{m=1}^{B}z_{im}\right)\left({z}_{jk}-\frac 1 B \sum\limits_{m=1}^{B}z_{jm}\right).
\end{equation}

The regression is performed by minimizing the $\chi^2$ ,which is defined as
\begin{equation}\label{eq:chi2}
\chi^2 = \left(\Phi_\mathrm{}(\mathbf{R}) - \mathbf{z} \right)^{\mathrm{T}} \bm{\Sigma}^{-1} \left(\Phi_\mathrm{}(\mathbf{R}) - \mathbf{z}\right),
\end{equation}
where $\mathbf{z} = \sum_{k=1}^{B}\mathbf{z}_k/B$ and $\Phi_\mathrm{}(\mathbf{R}) = (\Phi_\mathrm{}(R_1),...,\Phi_\mathrm{}(R_n))^\mathrm{T}$. Note that $\Phi$s are estimated from Eqn.~\ref{eq:phi} and they need to be in the same unit as $z$s (i.e. km/s).

\begin{figure}
\centering
\includegraphics[width=0.45\textwidth]{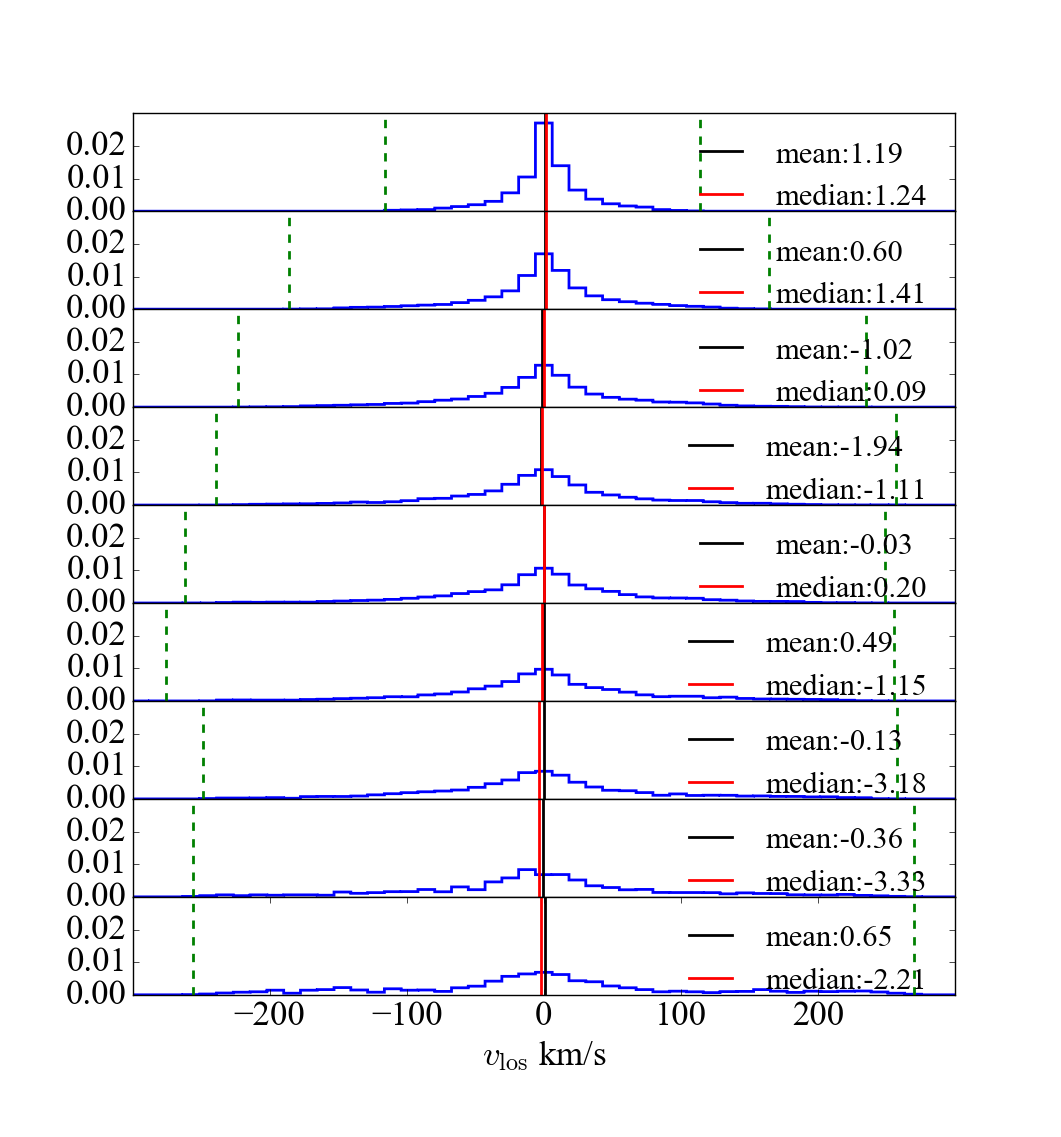}
\caption{The PDFs of velocities in radial bins. These bins are located at 0.75, 2.25, 3.75, 5.25, 6.75, 8.25, 9.75, 11.25, 12.75 kpc/$h$ from the top to the bottom panel. The black vertical lines show the mean and the red vertical lines show the median, with the numbers given in the legend of each subplot. Between two green dashed lines are the spaxels within $4\sigma$ to the median. The gravitational redshift signal will be seen as the relative shift of these distribution towards left from top to bottom. }
\label{fig:bin_distribution}
\end{figure}

The line-of-sight velocities of spaxels in radial bins after stacking these masked maps are shown in Fig.~\ref{fig:bin_distribution}. We find that the distributions are extremely noisy starting from the tenth bin so we discard that bin and those on larger scales. Also we discard the outliers whose velocities are farther than $4\sigma$ from the median (see green dashed lines in Fig.~\ref{fig:bin_distribution}). The spreads become larger as the radius increases. The PDFs are similar to Fig.~1 of \cite{wojtak2011}.

\begin{figure}
\centering
\includegraphics[width=.5\textwidth]{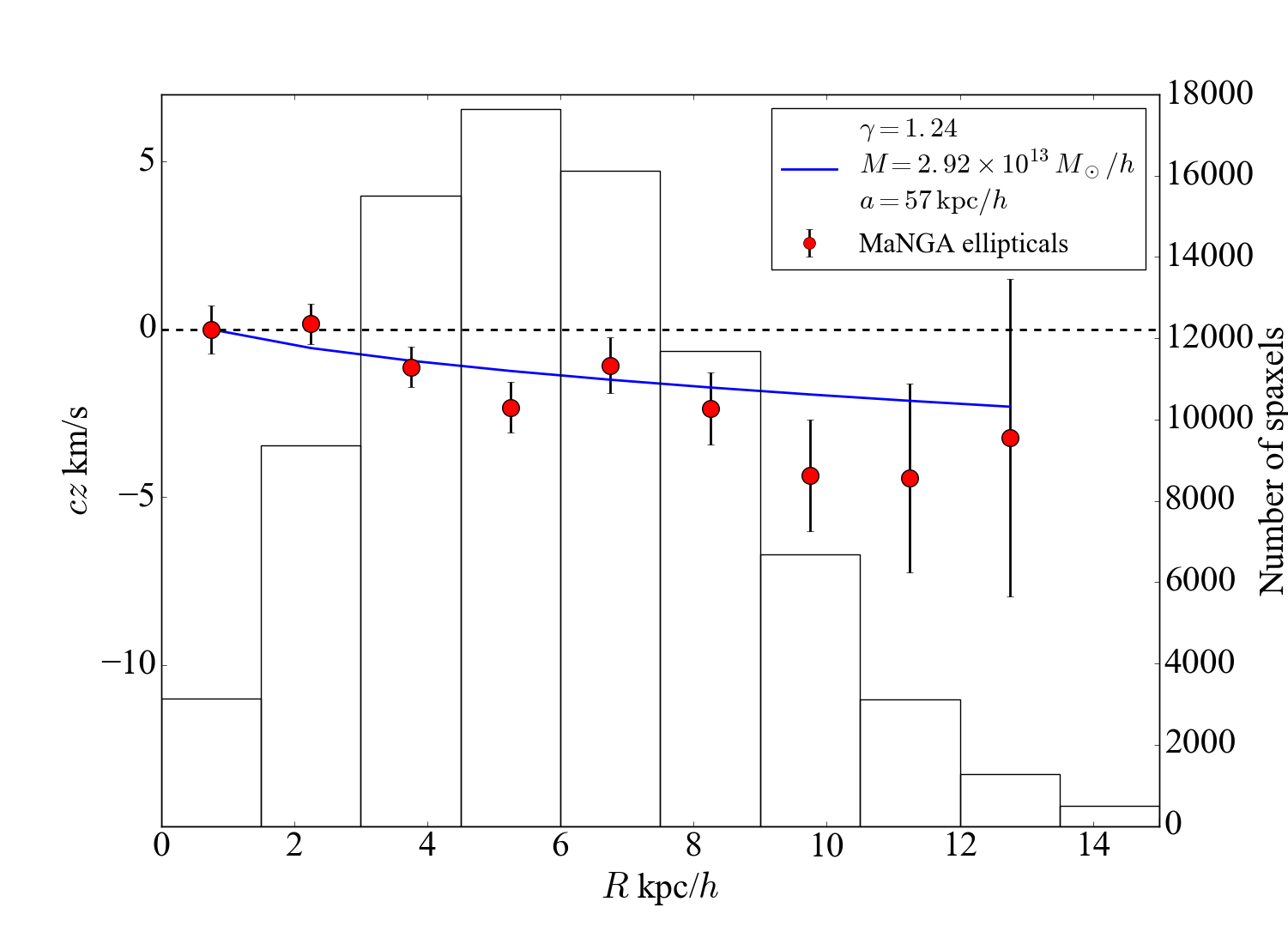}
\caption{Radial velocity distributions of galaxies combined from 272 MaNGA ellipticals. The line-of-sight velocities ($cz$) are plotted in bins of projected radius $R$ in physical scales kpc/$h$. The bar plot (right $y$-axis) shows numbers of spaxels in each radial bin. We can see the numbers increase as the radius increases first and then decrease as either not many galaxies have $R > 8$ kpc/$h$ or more bad spaxels are masked out. The red dots show the medians of line-of-sight velocities in each radial bin and the error bars are from 5000 bootstrapping samples of 272 ellipticals. The curve shows $\chi^2$ fitting of $\left(\Phi-\mathbf{v}^2/2\right)/c$ on the first 10 bins. The legend presents the best fitting parameters. We believe that this measurement, made using the standard MaNGA DAP pipeline, is contaminated by systematic effects, and therefore does not represent a measurement of the gravitational redshift profile.}
\label{fig:dap_fitting}
\end{figure}

Fig.~\ref{fig:dap_fitting} shows  the median of the stacked velocities and the numbers of spaxels in radial bins. We can see that there is a relative blueshift of a  few km/s from the outskirt to the center since the central potential is lower. 
The error bars get larger as the projected distance increases. The bar plot indicates the number of spaxels at different projected distances. As $R$ increases, the number increases first because the volumes of cylindrical shells increase and then decreases for two reasons. One is that the spaxels in the outskirts are more likely to be masked, and the other reason is that not many galaxies which have such large physical radii.

The numerical values of $z_{g}(r)$ show a steep gradient in the gravitational redshift, with a
relative redshift difference of $\sim 2-3$ km/s between the centers of the galaxies and points 10 kpc away. This is much larger than predictions on the basis of theoretical models. In order to investigate this further, we compute the best fit curve from $\left(\Phi-\mathbf{v}^2/2\right)/c$ where $\Phi$ is computed following Eqn.~\ref{eq:phi} as described in Sec.~\ref{sec:theory}. 
Marginalizing over the scale factor $a$ gives the $3\sigma$ confidence detection of a signal, with a one sigma logarithmic mass range from  $\log_{10}(M/M_\odot)=12.4$ to $\log_{10}(M/M_\odot)=13.3$. We expect that the inner regions of BCGs should be dominated by the gravitational effects of stars \citep{schmidt2007}, and so we would expect the mass inferred from the best-fitting gravitational
redshift profile to be close to that of the stellar mass. The sample mean stellar mass and
standard deviation is $\log_{10}(M/M_\odot)=11.03\pm0.15$, which is a factor of
$27$ times smaller than the measurement from $z_{g}$. 

As mentioned above, this redshift difference of 2-3 km/s is smaller than the accuracy for which the DAP pipeline results have been tested. It is therefore likely that systematic effects at this level are responsible for the signal that we see in Fig.~\ref{fig:dap_fitting}, and the results are not a measurement of gravitational redshift. In the next section we therefore carry out our own analysis of the raw MaNGA DRP outputs, using techniques designed to mitigate these systematics.

%% file: tex/dsp.tex
\section{Analysis of MaNGA DRP outputs} \label{sec:dsp}
\label{DRPintro}

As discussed in the previous section, the signal measured from the DAP outputs (Fig.~\ref{fig:dap_fitting} in Sec.~\ref{sec:dap}) is best fit with a mass $M=2.92\times 10^{13}$ $M_\odot/h$, which is much larger than expected ($M=1.07\times 10^{11}$ $M_\odot$ in Fig.~\ref{fig:sample_stats}).  There is a strong indication that the signal is probably not due to a gravitational redshift effect. At the level of a few km/s, there are a number of systematic effects which could be responsible. A possibility could be related to wavelength calibration (which has not been shown to be accurate below 5 km/s, \citealt{bundy2014, Yan2016}). The effect would need to systematically shift the wavelengths of the most luminous galaxies, or parts of galaxies with respect to their outer regions, or with respect to fainter galaxies. A CCD charge transfer effect could also be possible, which could  lead to a systematic error whereby stronger absorption lines are systematically offset relative to weaker ones. This effect has been seen in emission lines in comparison lamps where stronger lines are at slightly different apparent rest wavelengths from weaker ones, likely due to some charge transfer effect in the CCD. The inner parts of galaxies, where the continuum is stronger and absorption lines are also stronger due to higher metallicities, might be systematically shifted by such an effect relative to outer parts. 

We therefore believe that the standard MaNGA pipeline is not best suited for measuring the offsets of the order of 1 km/s or less which are produced by gravitational redshifts. Rather than attempting to recalibrate the standard pipeline or build a model for charge transfer effects \citep{Massey2014}, we instead have developed a data analysis technique which is specialized to measuring extremely small wavelength shifts from the raw data. We use cross-correlation techniques to find wavelength shifts between stacked fiber spectra. However, since the expected shifts are so small, we will be using an extremely small window (of the order of angstroms) to measure the cross-correlations. By design, therefore, we will be comparing parts of stacked spectra which are very similar in wavelength, and not too different in flux. Because of this, we  expect that any systematic effects due to charge transfer, wavelength calibration, or other unknown effects will be acting almost equally on both spectra in the comparison and therefore cancel out, except for any real physical effects such as gravitational redshift.
In this section we explain how we make these different relative velocity measurements from MaNGA DRP outputs. We contrast the results from those obtained using whole-spectrum template fitting provided in MaNGA DAP outputs (Sec.~\ref{sec:dap}).

\subsection{Stacked spectra}

\begin{figure*}
\centering
\includegraphics[width=1\textwidth]{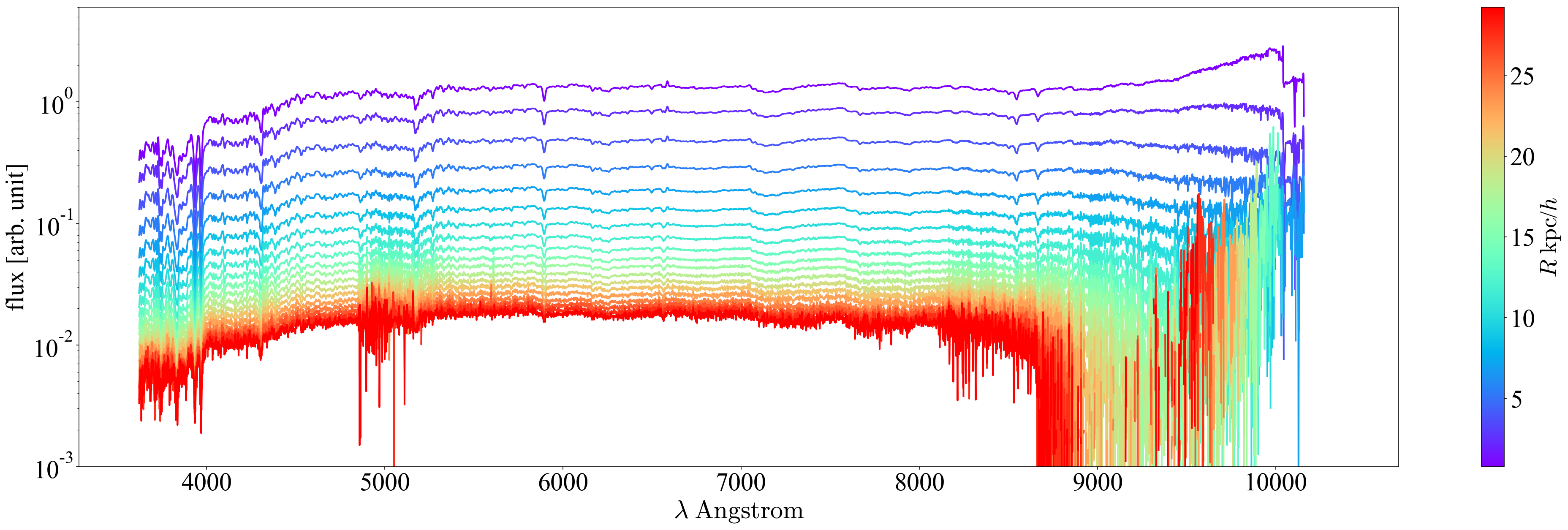}
\caption{The stacked spectra of 272 ellipticals. Spectra at different distances (0--30 kpc/$h$) from the centre of galaxy are shown in different colors. Note that the change in amplitude is not artificial but rather shows the real change in flux of galaxies with distance.}
\label{fig:dsp_stack}
\end{figure*}

Our first task is to generate a set of stacked spectra, where each stack only includes spectra taken within a certain range of physical distances from the galaxy center. We start with a MaNGA data cube object \texttt{LINCUBE} for each galaxy with 0.5 arcsec spatial pixels (NX$\times$NY spaxels), where each spaxel has linear wavelength sampling from 3622 to 10353 angstroms (spectral elements NWAVE=6732), for a total size of NX$\times$NY$\times$NWAVE pixels. The galaxy sample we use is described in Sec.~\ref{sec:data}. We first rescale lengths from arcsec to physical (kpc) scales. We then use interpolation to shift each spectrum to the rest frame wavelength by dividing the wavelength by $(1+z_\mathrm{center})$ where $z_\mathrm{center}$ is computed based on central velocities from the DAP outputs. This guarantees that the centers of galaxies are put into the rest frame for all galaxies. After that, satellite masks and mirror masks are applied as described in Sec.~\ref{sec:dap}. Stacking of the spectra is performed by averaging the flux in radial bins weighted by inverse variances (for our fiducial analysis). In order to make use of the approximately similar shapes of galaxies we scale the selected ellipticals so that their profiles are all in units of scale length $a$ and then stack. The stacked data in units of $R/a$ in this way are better matched and this should ensure that the stacking optimizes the signal rather than averaging together information from what are effectively different radii in different galaxies. We then report the results from averaging the flux in $R/a$ bins weighted by inverse variances. We also investigate the possible differences from this stack by making a stack with unit weights, in Sec.~\ref{sec:comparison}. The stacked spectra are shown in Fig.~\ref{fig:dsp_stack} where different radial bins are represented with different colors. We can see that spectra at inner radii have very clear and coherent characteristics while those at outer radii are much noisier. Based on theoretical models (Sec.~\ref{sec:theory}), the gravitational redshift corresponds to coherent shift of $\sim$ 0.01 angstrom at $\sim$5000 angstrom. We expect the spectra closer to the centers of galaxies (purple and blue lines in Fig.~\ref{fig:dsp_stack}) to be shifted systematically by this amount with respect to the spectra in the outskirts. This is obviously not visible from the plot, but can be constrained or maybe measured using cross-correlation techniques.

\subsection{Redshift profile}\label{sec:redshift_profile}

In order to make such small shift measurable, we cross-correlate pairs of the stacked spectra at different distances by shifting the one with a smaller distance in wavelength with a multiplication factor $\lambda_\mathrm{shift}$ and examining the the cross-correlation as a function of this shift. Mathematically, denoting as $f_i(\lambda)$ the flux in the $i$-th radial bin at the wavelength $\lambda$, the cross-correlation is performed between two series ($f_i(\lambda*\lambda_\mathrm{shift})$ and $f_j(\lambda)$ (assuming $i<j$, the $i$-th bin is closer to the center). The wavelength shift between the $i$-th and $j$-th bins is computed as
\begin{equation}\label{eq:cross}
    \lambda_{i,j}=\underset{\lambda_\mathrm{shift}}{\arg\max} \text{ Cross-correlate}(f_i(\lambda\cdot\lambda_\mathrm{shift}), f_j(\lambda)),
\end{equation}
where the redshift $z_{i,j}=\lambda_{i,j}-1$.
We use the central bin (0th bin) for reference and cross-correlate all the other radial stacked spectra with $f_0(\lambda)$ to obtain $z_{0,j}$ for all the $j$s. Note that $z_{0,0}$ is precisely set to 0 as a reference zero point. 

We perform this computation using the data over the range from 3622 to 8100 angstroms, where most of the spectral shape characteristics and template absorption lines are included. 
The cross-correlation is only performed over a finite and small window in wavelength, however,
for the reasons mentioned above, in Sec.~\ref{DRPintro}. We vary this window size in order to check how small it should be to avoid systematic effects.
Different window sizes split the whole range into different numbers of windows where the cross-correlations are performed. For example, the window size $\Delta\lambda=100$ angstrom would split the range 3622 to 8100 angstroms into 44 windows, 3622 to 3722, 3722 to 3822,..., 7922 to 8022 angstroms. 44 $z_{0,j}$s are obtained from these windows. The median and covariance are further computed from 44 $z_{0,j}$s in each radial bin. We use the median rather than the mean shift in order to be less sensitive to outliers.

\begin{figure}
\centering
\includegraphics[width=1\columnwidth]{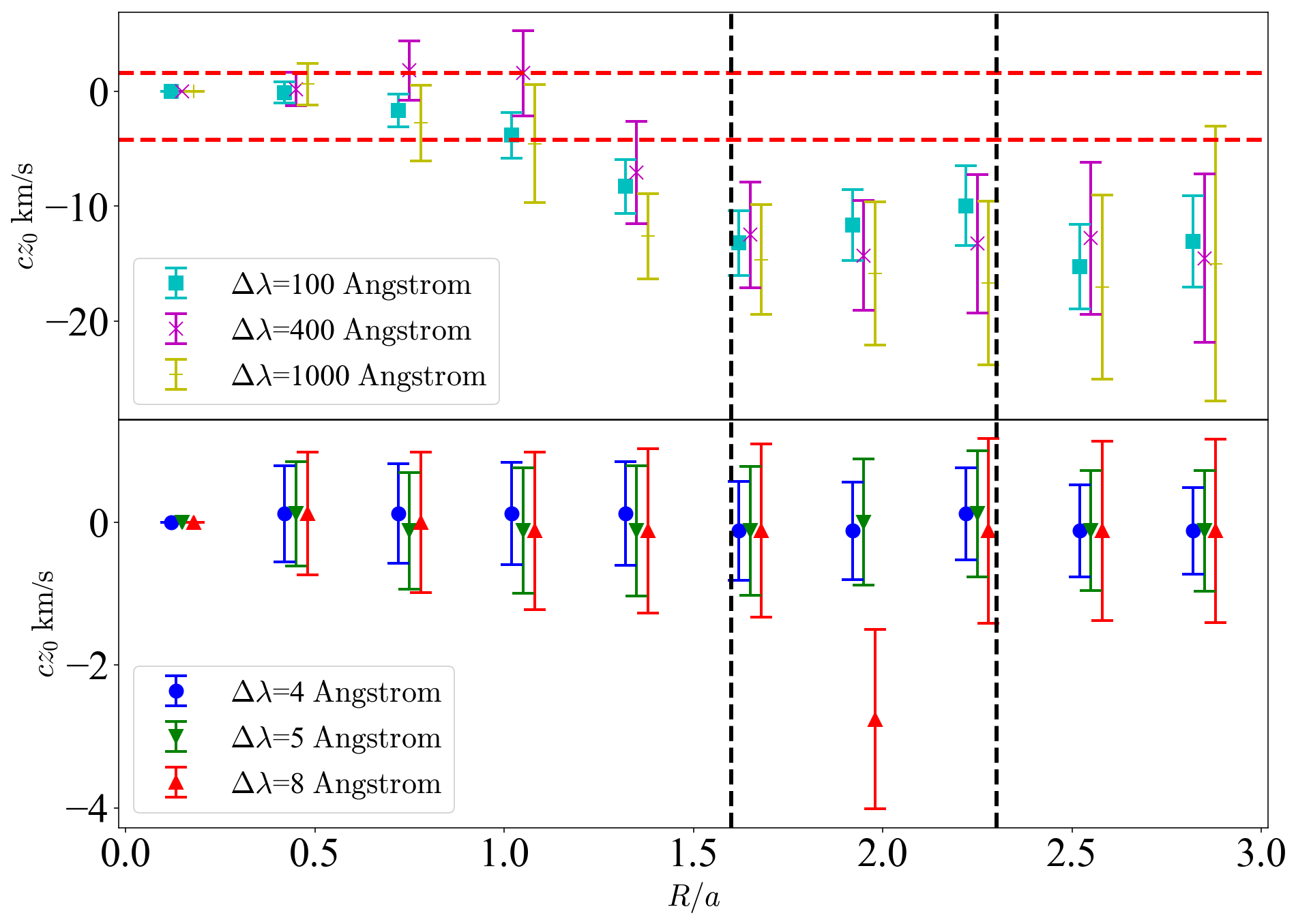}
\caption{The spectra redshift $z_{0,j}$ in bins of projected radius scaled by scale length in $R/a$ of stacked ellipticals. The panels are splitted according to large and small window sizes $\Delta\lambda$. The top panel shows the spectra redshift with $\Delta\lambda=100, 400, 1000$ angstrom. The two horizontal dashed red lines indicate the $y$-axis limits of the bottom panel. The region between two vertical dashed black lines is used for the computation of $z_{0,j}(1.6<R/a<2.3)$ in Fig.~\ref{fig:convergence}. The bottom panel shows the spectra redshift with $\Delta\lambda=4, 5, 8$ angstrom. The error bars (errors on the median) are computed as 1.2533 times errors on the mean.}
\label{fig:dsp_window}
\end{figure}

\begin{figure}
\centering
\includegraphics[width=1\columnwidth]{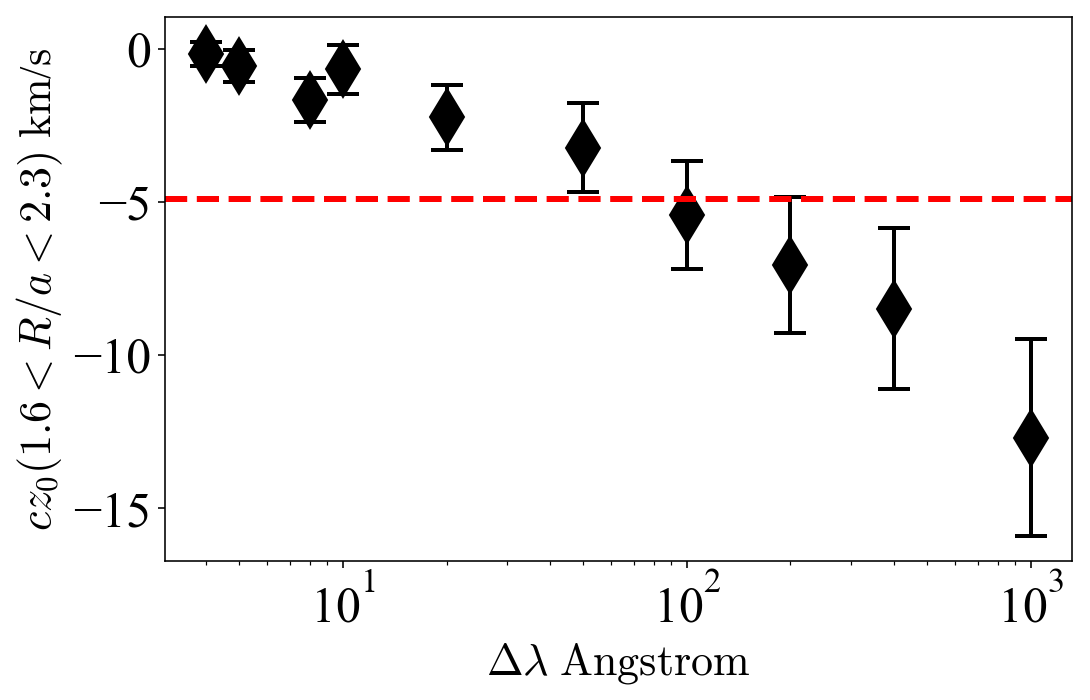}
\caption{The mean spectra redshift $z_{0,j}$ between $R=1.6a$ and $R=2.3a$ (between black dashed line in Fig.~\ref{fig:dsp_window}). The error bars (errors on the median) are computed as 1.2533 times errors on the mean. The red dashed line shows the same quantity $z_{0,j}(1.6<R/a<2.3)$ from MaNGA DAP for comparison.}
\label{fig:convergence}
\end{figure}

 We find that the signal $z_{0,j}$ is indeed sensitive to the window size $\Delta\lambda$ (see Fig.~\ref{fig:dsp_window}). We find that with a window size of 50 angtroms or more, the wavelength shifts are large, comparable to those seen using the MaNGA DAP data. This can be seen in the top panel of Fig.~\ref{fig:dsp_window}. In this case, as with MaNGA DAP, the velocities are obtained by using the information from widely separated absorption lines and other spectral features. When the wavelength window is reduced to 10 Angstroms or less, the wavelength shifts converge  to smaller values. This can be seen in the bottom panel of Fig.~\ref{fig:dsp_window}, where we show results for smaller window sizes. In this case the wavelength shifts are much smaller, around 1 km/s or less, comparable to the expected gravitational redshift effect. The error bar sizes are smaller for the smaller windows, as there are more of them to average over. We further illustrate the convergence of $z_{0,j}$ with cross-correlation window size by comparing values of $z_{0,j}$ at $R\sim R_e$ ($R\sim 2a$) in Fig.~\ref{fig:convergence} as a function of $\Delta \lambda$ The plot is consistent with what is seen in Fig.~\ref{fig:dsp_window}.  Our median gravitational redshift profile results are insensitive to the choice of $\Delta \lambda$ as long as $\Delta \lambda < 10$ Angstroms.

\subsection{Model fit}

\begin{figure}
\centering
\includegraphics[width=1\columnwidth]{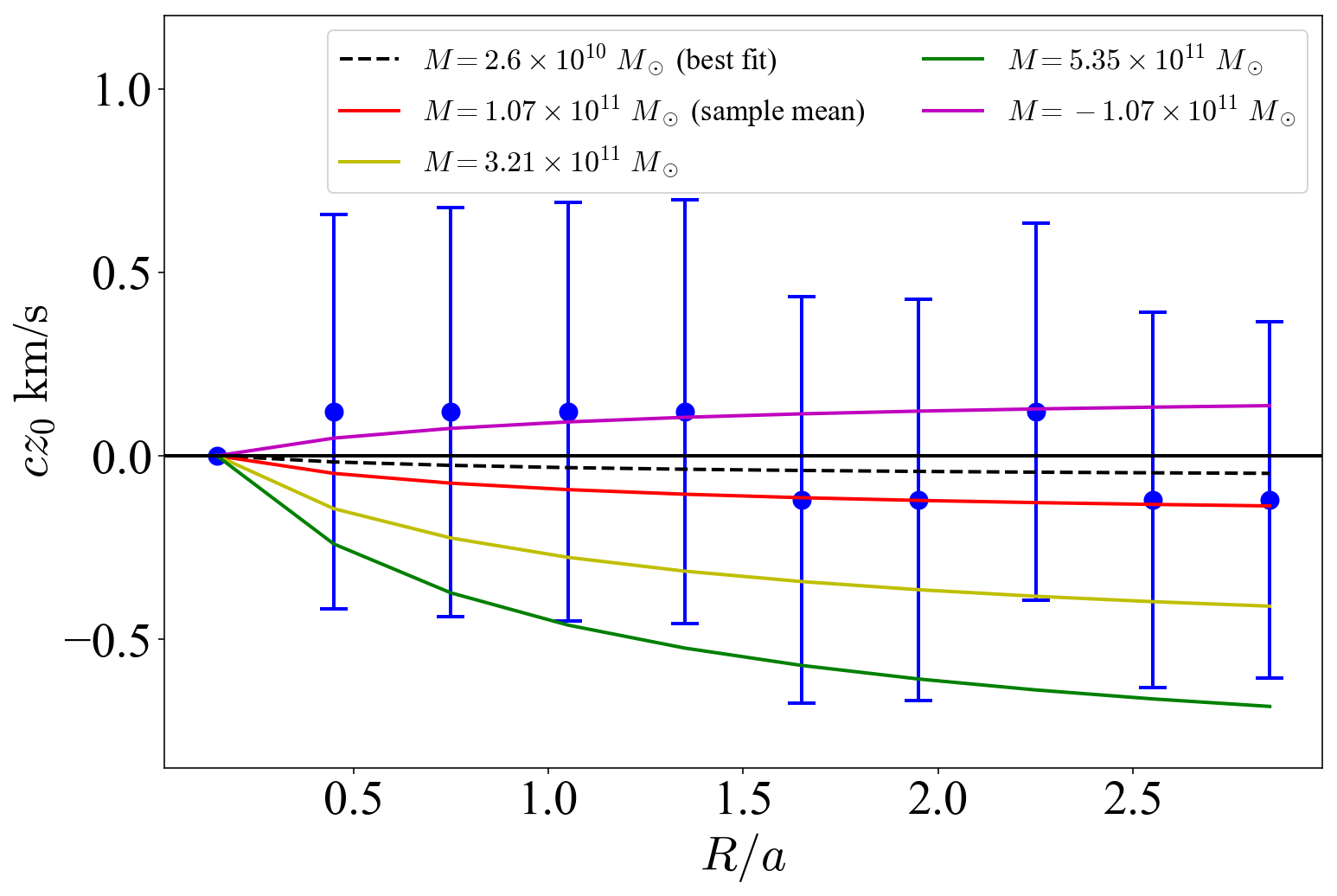}
\caption{$\chi^2$ fittings to the spectra redshift with $\Delta\lambda=4$ angstrom. The black dashed line shows the theoretical predictions estimated from the best fitting stellar mass. The red, yellow, green and magenta solid lines represent the sample mean stellar mass, 5, 10 multiples and also the negative of the sample mean respectively. }
\label{fig:dsp_fitting}
\end{figure}

We therefore proceed to fit the wavelength shift signal of the smallest window size $\Delta\lambda=4$ angstrom using the equations Eqn.~\ref{eq:jhmodels} and Eqn.~\ref{eq:phi} from Sec.~\ref{sec:theory} under the constraint that the center bin is a fixed zero point As discussed in Sec.~\ref{sec:dap} the transverse Doppler effect is so small compared to gravitational redshift effect that it is a good assumption to ignore it and  only fit $\Phi$ to the data.
The data points are shown in Fig.~\ref{fig:dsp_fitting}, along with the best fit curve and the curves which are
predictions for the profile due to the stellar mass. We can see that the redshift difference between the stacked galaxy center and the outskirts at radius
$R/a=2$ is of the order of 0.01 km/s, but the error bars are large, so that, as we show below, larger redshift differences, of order of 0.5 km/s, are also consistent with the measurement. The error bars are too large to yield any significant detection of
a gravitational redshift.

\begin{figure}
\centering
\includegraphics[width=1\columnwidth]{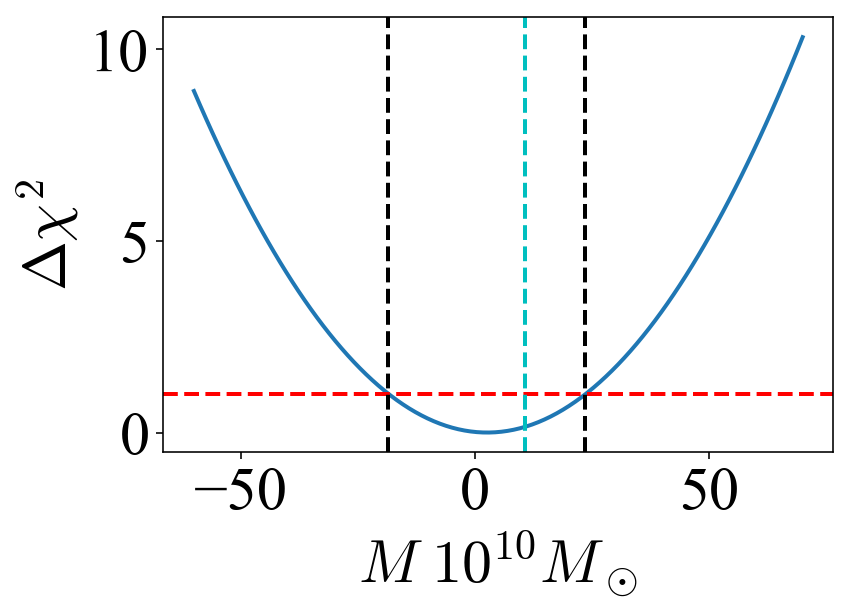}
\caption{$\Delta\chi^2$ function of the fit. The region between two black dashed lines shows a 68.3\% confidence interval. The vertical cyan dashed line is the mean of the stellar mass in the sample.}
\label{fig:dsp_chi2}
\end{figure}

We carry out fits using the shape of the stellar mass profiles as a constraint. We fix the scale radius to be the mean stellar scale radius of the sample measured from the
galaxy images. The covariance matrix is evaluated based on the wavelength shifts from the different $R/a$ bins. If we recall that the sample mean stellar mass is $\log_{10}(M/M_\odot)=11.03\pm0.15$ or $M=1.07\times10^{11}$ $M_\odot$/$h$ this means that an elliptical galaxy dominated by the stellar component should have a $z_{g}$ profile consistent with this value. From our mass fitting, we find the best fit total galaxy mass measured from the gravitational redshift is $M=(2.6\pm21.0)\times10^{10}$ $M_\odot$/$h$. We  show theoretical predictions for different masses (including a negative mass) in Fig.~\ref{fig:dsp_fitting}. The $\Delta\chi^2$ curve as a function of mass is shown in Fig.~\ref{fig:dsp_chi2}. We can see that the sample mean stellar mass is within the the 68.3\% significance contours and thus within $1\sigma$ of fitted parameters.
Subtracting the mean stellar mass from the 1$\sigma$ upper boundary on the fitted mass, we find that the difference (the dark matter mass) is restricted to
be $M_\mathrm{dm}< 1.3\times10^{11}$ $M_\odot$/$h$, or less than $1.2$ times the stellar mass.

\subsection{Robustness of the Measurements} \label{sec:comparison}

\begin{figure}
\centering
\includegraphics[width=1\columnwidth]{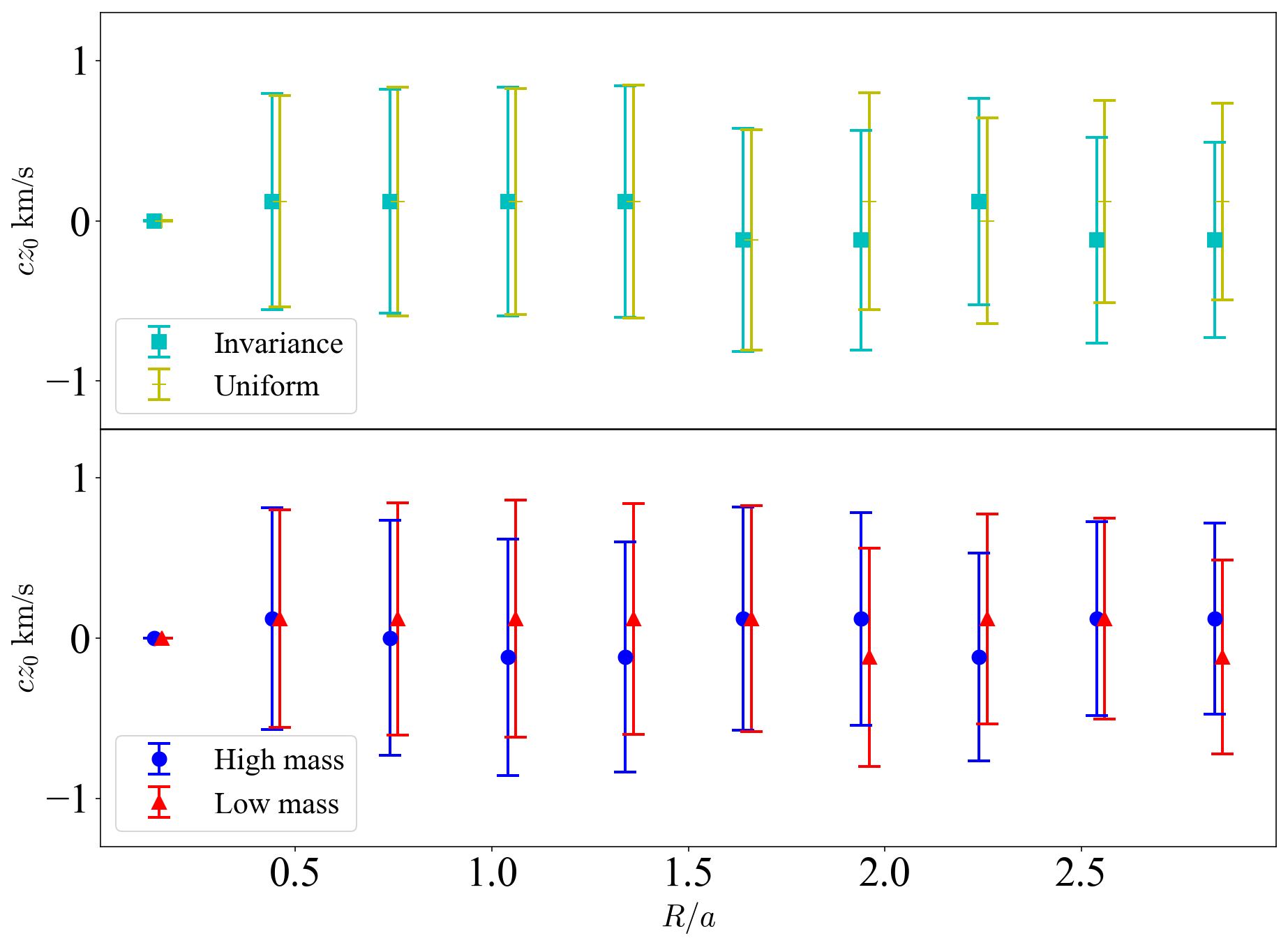}
\caption{The top and bottom panels show the comparison between different weighting schemes and the test between high and low mass galaxies respectively. $z_{0,j}$s are computed based on $\Delta\lambda=4$ angstrom.}
\label{fig:comparison}
\end{figure}

As tests of the robustness of our measurement, We explore a different weighting scheme, uniform weighting as opposed to inverse variance weighting. Although the inverse variance weighting usually optimizes stacked variances to provide signals at a high significance level, it relies heavily on those ``well detected'' galaxies and thus may be very sensitive to a small number of objects. The uniform weighting, on the other hand, compensates for such an issue and comparing one weighting scheme to the other effectively covers the whole range of reasonable weighting schemes. From the top panel in Fig.~\ref{fig:comparison}, we see that $z_{0,j}$s are  very similar for the two different schemes. Therefore, the choice of weighting schemes is not a significant factor in the signal. 

We also split the whole sample by mass to create high and low mass samples. If we had a well measured gravitational redshift signal, this test would be useful to help confirm the reality of the effect. In our case, the test serves to highlight any anomalous results. If the signal were dominated by the gravitational redshift effect, then in $z_{0,j}$s a gap would be expected between the results from high mass and low mass samples. From the bottom panel in Fig.~\ref{fig:comparison}, both sets of data give very similar results, as we expect given the large error bars. 


\subsection{Window correlation test}\label{sec:window_test}
Since we split the whole spectra into small windows and cross-correlate those, it is not trivial to determine if the above methodology maintains the same statistics compared to shifting and cross-correlating the whole spectra. It is possible that cross-correlating spectra based on many separate windows introduces a potential bias that is able to systematically change the signals from different window sizes, leading to what we have seen in Fig.~\ref{fig:dsp_window}. We therefore conduct ``window correlation tests'' where we apply known wavelength shifts by hand to the spectra and see how well our cross-correlation analyses can recover them.

\begin{figure}
\centering
\includegraphics[width=1\columnwidth]{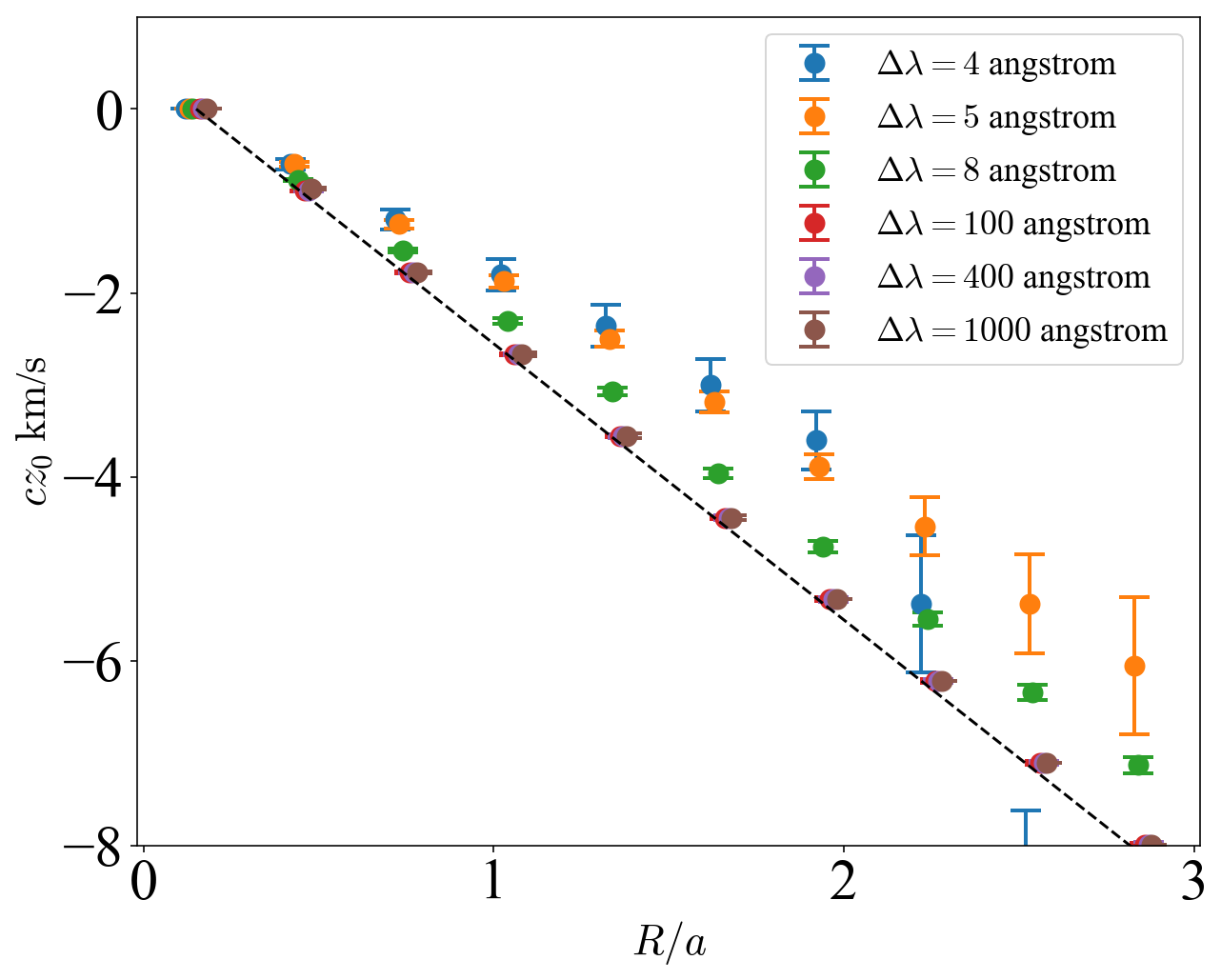}
\caption{The spectra redshift $z_{0,j}$ in bins of projected radius scaled by scale length in $R/a$ of stacked ellipticals from the window correlation test as described in Sec.~\ref{sec:window_test} with $\Delta_\lambda=4,5,8,100,400,1000$. The dots represent the medians and the error bars (errors on the median) are computed as 1.2533 times errors on the mean. The black dashed line shows the expected theoretical result.}
\label{fig:window_test}
\end{figure}

We take the stacked spectra of the 1st bin as seen in Fig.~\ref{fig:dsp_stack} and shift each bin from the last by multiplying by   $z_\mathrm{shift}=1.000003$. This gives a 0.9 km/s redshift between any two consecutive bins. Then as we did in Sec.~\ref{sec:redshift_profile} we split the range 3622 to 8100 angstroms into different windows whose sizes are $\Delta_\lambda=4,5,8,100,400,1000$ respectively. For each splitting scheme, we cross-correlate the split spectra in the 2nd, 3rd bin etc. with the 1st bin. Fig.~\ref{fig:window_test} shows the $z_{0,j}$ as a function of $R/a$. We find that cross-correlating on larger window sizes $\Delta_\lambda=100,400,1000$ is able to almost fully recover the redshift compared to the expected curve shown by the black dashed line. We also find that cross-correlating with smaller window sizes $\Delta_\lambda=4,5,8$ can recover the shift within a certain percentage error (about 25\% from Fig.~\ref{fig:window_test}). This systematic error of up to $25\%$ is much smaller that the relative difference between  cross-correlating with large and small window sizes in Fig.~\ref{fig:dsp_stack}. Because of this and the fact that the 25\% is much smaller than the statistical error on the gravitational redshift measurement, we therefore find that the split window cross-correlation technique doesn't introduce any significant bias in the measurements.

%% file: tex/discussion.tex
\section{Discussion}\label{sec:discussion}

The expected amplitude of the gravitational redshift is much below the resolution limit of the spectrograph. As a result different systematic uncertainties such as wavelength calibration problems and charge transfer may come to lower the detectability. Although we are unable to  make a significant detection of a signal, we find convergence with different data window sizes used in our analysis, and successful recovery of small test wavelength shifts inserted by hand. We therefore  believe we have been successful in filtering out the systematic uncertainties mentioned above and provide most precise measurement of such redshifts at these scale.

Other current dark matter probes also suffer from systematic uncertainties. For example, weak lensing has been a powerful cosmological probe of the matter density in galaxies and galaxy clusters. It is not only affected by uncertainties such as shear calibration errors, poorly determined background source redshifts  and source obscuration but also the effects of incorrect assumptions about cluster centering, halo triaxiality and projection effects \citep{Mandelbaum2016, Mandelbaum2016b}. Another probe, X-ray observations, assumes hydrostatic equilibrium \citep{Schindler1996, schmidt2007}, which is not strictly valid under most circumstances. The gravitational redshift allows us to probe mass directly without projection effects
and avoiding the specific systematic uncertainties of these other methods.

 Measurements of gravitational redshift on scales of few Mpc around clusters \citep{wojtak2011, sadeh2015} and using galaxy cross-correlation \citep{Alam2016Measurement}  have been conducted relatively recently and have been able to detect the signal with   $2-3\sigma$ significance. These constraints on gravitational redshift are consistent with the predictions of models of gravity and therefore have been used to constrain models such as Modified Newtonian Dynamics (MOND). The smaller scale measurements in this paper, are able to help us understand the inner profile of galaxies as well as the non-linear regime and baryonic physics. Better precision will be obtainable in the future.

Future surveys with better samples (high Charge Transfer Efficiency and reduced number of transactions \citealt{PASP}, better wavelength calibration, etc.) of massive spatially-resolved galaxies will be capable of detecting gravitational redshift signals, likely at a reasonable significance level. For example the ongoing MaNGA project \citep{bundy2014} will include eventually include  $\sim250$ BCGs. Also the MASSIVE Survey \citep{Ma2014} which currently probes $\sim$100 most massive early-type galaxies known with stellar mass $M>10^{11.5}\,M_\odot$ makes a data set which can be used already to attempt such a measurement. Furthermore, as is pointed out in \cite{coggins2003}, the degree of central concentration in the galaxy is a very important factor. Highly concentrated galaxies might be even better targets than massive ellipticals for optimizing the detectability of the gravitational redshift effect within galaxies. It could be worth making a large sample of those galaxies such as compact dwarf galaxies. For example, the predicted signal from the dwarf galaxy M32 (velocity dispersion $\sigma$ $\sim100$ km/s) is larger than a typical giant elliptical with $\sigma$ $\sim300$ km/s. The reasons are the following. Firstly, a highly concentrated galaxy increases the size of the signal due to its particularly deep central potential well. Secondly, as more of the light originates from a more centrally-concentrated system, integration along the line of sight leads to a less diluted signal.

%% file: tex/conclusion.tex
\section{Conclusion}\label{sec:conclusion}

The gravitational redshift effect allows one to directly probe gravity and is independent of cosmology. One of the applications is to put constraints on galaxy masses, for example on the dark matter content of elliptical galaxies. The gravitational redshift is a relatively condition-free and assumption-free probe at such small scale compared to lensing (which is subject to mass-sheet degeneracies and modeling uncertainties) or stellar dynamics. In this paper, we have explored measurements of gravitational redshifts in elliptical galaxies using two MaNGA data products, DAP and DRP. We find that using standard velocity pipeline (DAP) results in what is likely a spurious signal, much larger in amplitude than expected from gravitational redshifts. As the DAP results were not designed to measure $\sim$1 km/s wavelength shifts, and have not been tested for accuracy at these levels, it is reasonable to believe that systematic effects from charge-transfer, wavelength calibration residuals or other unknown effects could be the cause. We have therefore developed a new analysis method on the raw data (DRP products), breaking spectra into small windows, of order a few angstroms in length, and only measuring the wavelength shifts (from cross-correlation techniques) using the data in each window. We found that below 10 angstroms the redshift profiles of elliptical galaxies so obtained converge to
results different to (and much smaller) than those obtained using the standard (DAP) pipeline.

From our DRP analysis we find no significant evidence of a gravitational redshift signal
from the stacked sample of 272 elliptical galaxies. Fitting theoretical galaxy models we do however find 
a constraint on the total mass to be $2.6\pm21.0\times10^{10}$ $M_\odot$, which is consistent with the mean stellar mass 
$M=10.7\times10^{10}$ $M_\odot$. The dark matter content of BCGs is therefore constrained within radius 15 kpc/$h$ to be less than $1.2$ times the stellar mass, at the 1$\sigma$ level. Given how close the measurements are 
to a detection, it should be possible to make
a significant measurement of the gravitational redshift with a modestly larger sample. In the future, much larger samples of IFU galaxy spectra may become available. For example, with the HECTOR survey \citep{2015IAUS..309...21B} that is planned to include integral field spectroscopy of 100,000 galaxies, relativistic effects could be studied in detail and may become important probes of gravity and the dark matter content of galaxies.

%% file: tex/appendix.tex
\section*{Appendix}